\documentclass[epj,twocolumn]{webofc}
\usepackage[varg]{txfonts}   
\usepackage{amsmath}

\def\EmissT{\,{}/ \hspace{-1.5ex}  E_{T}}
\newcommand{\lsim}{\stackrel{<}{_\sim}}

\newcommand{\gev}{\, {\rm GeV}}

\woctitle{LHCP 2013}
\begin{document}
\title{Squark flavour violation and naturalness at the LHC}
%
%

\author{Monika Blanke\inst{1}\fnsep\thanks{\email{monika.blanke@cern.ch}} 
}

\institute{CERN Theory Division, CH-1211 Geneva 23, Switzerland}

\abstract{%
While rare meson decays place stringent constraints on much of the flavour violating parameter space of supersymmetric models, the mixing between right-handed top and charm squarks is so far unconstrained by the available data. Such mixing has in fact a very appealing phenomenology: it significantly weakens the current experimental bounds from direct squark searches and leads to a modest reduction of fine-tuning. The scenario can be tested already with the present 8\,TeV data set by looking for the flavour violating final state $t\bar c(c\bar t)+\EmissT$, thanks to the large squark pair production cross section.
}
\maketitle
\section{Introduction}

After the discovery of the Higgs boson~\cite{:2012gk,:2012gu}, one of the main questions of particle physics to be answered is the stabilisation of the electroweak (EW) symmetry breaking scale. In the Standard Model (SM) large radiative corrections lead to a quadratic dependence of the Higgs squared mass  parameter on the cut-off scale $\Lambda$, and thus, in order to keep the EW scale at $\mathcal{O}(100\,\text{GeV})$, to an unnatural fine-tuning of parameters.

Many theoretical ideas have been put forward to solve the naturalness problem of the SM by introducing a new symmetry protecting the Higgs mass. Among those, supersymmetry has attracted by far the most attention both by theoretical considerations and experimental searches. In this case the scalar partners of the top quarks, the stops, cancel the quadratic divergences arising from top loops. 
The remaining leading-log one-loop contribution to the Higgs mass parameter $m_{H_u}^2$ reads
\begin{equation}\label{eq:deltamHu}
\delta m_{Hu}^2=-\frac{3Y_t^2}{8\pi^2}\left(m_{\tilde t_1}^2+m_{\tilde t_2}^2+|A_t|^2\right)\log\frac{\Lambda}{m_{\tilde t}}\,,
\end{equation}
so that the stop masses have to be below the TeV scale in order to preserve naturalness. Note that the equality of top and stop couplings to the Higgs field, given by the Yukawa coupling $Y_t$, is crucial for the cancellation of quadratic divergences. This relation gives rise to a testable relation of squark masses and mixing angles, the SUSY-Yukawa sum rule \cite{Blanke:2010cm}.

\begin{figure}
\centering
\includegraphics[width=.475\textwidth]{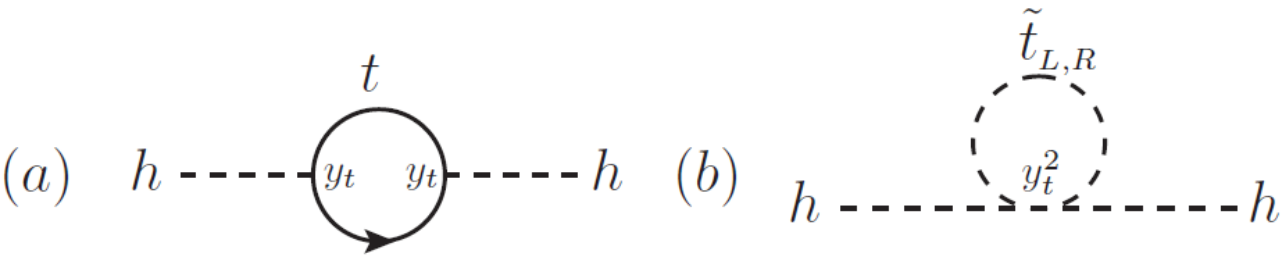}
\caption{(a) One-loop quadratically divergent contribution to the Higgs mass squared parameter from top quark loops. (b) Supersymmetric counterpart with top squarks that cancels the quadratic divergence in (a). \label{fig:loop}}
\end{figure}

Direct stop searches at ATLAS and CMS \cite{stop-search-status} have so far been unsuccessful and place increasingly stringent bounds on the mass of these particles. Therefore a lot of effort has been put into the identification and exploration of regions of parameter space where the stops might be hiding from direct observation.
In this spirit in \cite{Blanke:2013uia} we explored the impact of squark flavour mixing on the stop phenomenology at the LHC and the fine-tuning in the EW sector. A similar approach has recently been taken in \cite{Agrawal:2013kha}.

Flavour mixing in the down squark sector -- and due to the $SU(2)_L$ symmetry also the left-handed up squark sector -- is severely constrained by the precise measurements of $K$ and $B$ meson decays, see e.\,g.\ \cite{Giudice:2008uk,Gedalia:2010mf} for details. On the other hand relevant constraints on right-handed up sector flavour violation are available only from $D$ physics.  Consequently the mixing between the right-handed stop with either the right-handed up or scharm quark is left-unconstrained, as long as the product of both mixing angles is kept small.

Since the bounds on a second generation squark are much weaker than the ones for the first generation \cite{Mahbubani:2012qq}, we analyse the impact of a large mixing angle $\theta_R^{ct}$ between the right-handed stop and scharm.\footnote{For earlier related studies, see e.\,g.\ \cite{Han:2003qe,
Cao:2006xb,
Hiller:2008wp,
Kribs:2009zy,
Hurth:2009ke,
Bruhnke:2010rh,
Bartl:2010du,Bartl:2012tx}.} For simplicity we restrict ourselves to the case of a massless and mostly bino LSP, and assume the trilinear coupling $A_t$ to be small. The squared mass of the right-handed stop in eq.\ \eqref{eq:deltamHu} is then replaced by $c^2 m_1^2 +s^2 m_2^2$, where $c=\cos\theta_R^{ct}$, $s=\sin\theta_R^{ct}$, and $m_1$ and $m_2$ are the masses of the stop- and scharm-like states respectively.

In this context let us note that the stop mass bounds obtained from final states with one or two leptons place stronger bounds on scenarios with right-handed tops in the final state than on their left-handed counterparts. The origin of this asymmetry lies in the parity violation of weak interactions, resulting in a harder lepton $p_T$ spectrum in the case of right-handed tops~\cite{Perelstein:2008zt,Gedalia:2009ym,Belanger:2012tm,Almeida:2008tp,Rehermann:2010vq}.\footnote{Note that the recent ATLAS and CMS 1-lepton analyses \cite{CMS-PAS-SUS-13-011,ATLAS-SUSY-2013-037} explicitly study the stop mass bounds for  both top quark chiralities.} Consequently assuming the lightest neutralino to be mostly gaugino, bounds from stop searches with final state leptons constrain mostly the right-handed stop. It is therefore of particular interest to study the impact of flavour violation on these bounds.

\section{Bounds on the mixed squark masses}

\begin{figure}
\centering
\includegraphics[width=.475\textwidth]{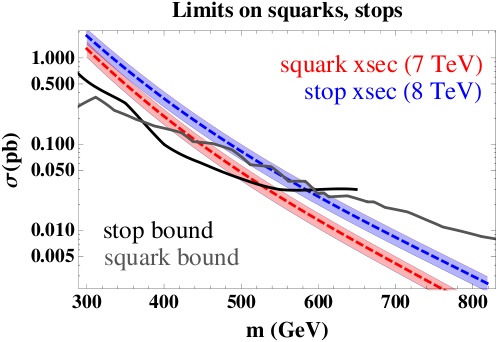}
\caption{Various upper limits on squark pair-production cross sections. In gray we show the envelope of experimental bounds on right-handed scharm for a massless neutralino, from $7\,{\rm TeV}$ $5\,{\rm fb}^{-1}$ ATLAS and CMS $\text{jets}+\EmissT$ searches~\cite{ATLAS:2012ona, :2012mfa, CMS:2012yua,Chatrchyan:2012wa, Aad:2012fqa}. In black we show the corresponding limit on stop pair production taken from~\cite{ATLAS-CONF-2012-166}. The red (blue) band corresponds to the theory  prediction for the scharm (stop) pair production~\cite{Beenakker:2011fu}. \label{fig:xsec}}
\end{figure}

In order to estimate the bounds on the masses of the mixed scharm-stop states, we reinterpret the constraints on the $t\bar t+\EmissT$ and $\text{jets}+\EmissT$ final states as summarized in Figure~\ref{fig:xsec}. 

In presence of a sizeable mixing angle the cross sections for squark pair production with subsequent decay into $t\bar t+\EmissT$ or $c\bar c +\EmissT$ get dressed by factors of $c^4$ and $s^4$. Since both squark mass eigenstates contribute to both experimental final states, we are led to
construct the following \ $\chi^2$ function to estimate the constraints on the masses and mixing angle of the mixed $\tilde t_R -\tilde c_R$ system:
\begin{equation}\label{eq:chi2}
\chi^2=\left[\frac{c^4 \sigma(m_1)+r_{t\bar t} s^4 \sigma(m_2)}{\Delta \sigma_{t\bar t}(m_1)}\right]^2+
\left[\frac{A \sigma(m_1)+r_\text{jets} B \sigma(m_2)}{\Delta \sigma_\text{jets}(m_1)}\right]^2
\end{equation}
Here we deduce the $1\sigma$ experimental uncertainty $\Delta \sigma_f(m_1)$  from the measured 95\%CL bounds assuming Gaussian errors and zero mean. Furthermore the factor $r_f\equiv \Delta \sigma_f(m_1)/\Delta \sigma_f(m_2)$ (for $f=t\bar t$, jets) is the ratio of experimental sensitivities to squarks with masses $m_1$ and $m_2$  and is a naive guess for how much the higher mass squarks contribute to the analysis.

\begin{figure}
\centering
\includegraphics[width=.475\textwidth]{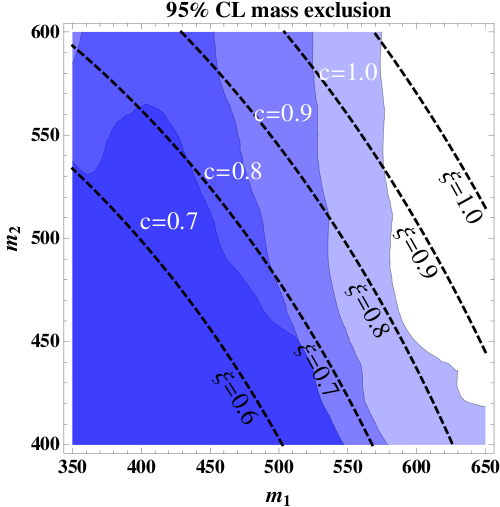}
\caption{Estimated bounds \cite{Blanke:2013uia} on the mostly-stop ($m_1$) and mostly-scharm ($m_2$) masses for various values of the mixing parameter $c=\cos\theta_R^{ct}$ as indicated, obtained in the aggressive  approach. The value of the naturalness parameter $\xi$, obtained by setting $c=0.7$, is indicated by the dashed lines.\label{fig:bound-agg}}
\end{figure}

As a naive guess, we can set $A=s^4$ and $B=c^4$ in \eqref{eq:chi2}. This approach completely neglects the mixed $t\bar c(c\bar t)+\EmissT$ final state which is unavoidably generated in the presence of a non-zero mixing angle $\theta_R^{ct}$. We therefore refer to this approach as the {\it aggressive} one, as it clearly underestimates the exclusion limits on the squark masses. Figure \ref{fig:bound-agg} shows the result of our naive $\chi^2$ fit in the $(m_1,m_2)$ plane, for various values of the mixing parameter $c$. We observe that compared to the lower bound $m_{\tilde t_R}>585\,\text{GeV}$ for flavour conserving stops \cite{ATLAS-CONF-2012-166}, the stop-like mass can be reduced significantly if the mixing angle is large.

\begin{figure}
\centering
\includegraphics[width=.475\textwidth]{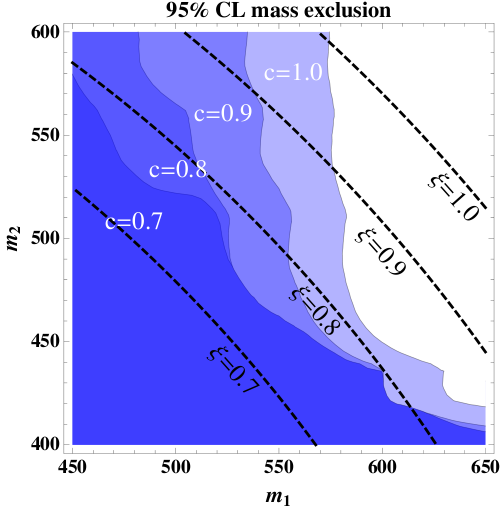}
\caption{Same as Figure~\ref{fig:bound-agg}, but for the conservative approach.\label{fig:bound-cons}}
\end{figure}

In reality even though no dedicated search exists the $t\bar c(c\bar t)+\EmissT$ final state is already being constrained by the $\text{jets}+\EmissT$ searches. We approximate the contribution of this flavour violating signal to the total $\text{jets}+\EmissT$ cross-section by assuming that it contributes in the same way as the $c\bar c+\EmissT$ final state if the top decays hadronically. We then have $A=s^4+2s^2c^2 Br(W\to jj)$ and $B=c^4+2s^2c^2 Br(W\to jj)$. Since in practice the four jet final state in this case looks different from the two jet case of flavour conserving decays, we expect this approach to yield a rather {\it conservative} estimate of the true situation. From Figure \ref{fig:bound-cons} it is clear that while the bounds in the conservative approach are stronger than in the aggressive approach, in particular for large mixing angles, a considerable improvement with respect to the flavour conserving case is still possible.

In order to quantify the improvement obtained for the fine-tuning in the Higgs mass parameter $m_{H_u}^2$, we define the tuning parameter
\begin{equation}\label{eq:xi}
\xi=\frac{c^2 m_1^2 + s^2 m_2^2}{m_0^2},
\end{equation}
where $m_0=585$ GeV is the experimental bound on the right-handed stop mass without mixing obtained in \cite{ATLAS-CONF-2012-166}. $\xi$ parametrises the change in the one loop contribution $\delta m_{H_u}^2$ generated by the right-handed stop-scharm mixing. Contours of constant $\xi$, evaluated for $c=0.7$, have been overlaid in Figures \ref{fig:bound-agg} and \ref{fig:bound-cons} for illustrative purposes. A mild improvement of naturalness, i.\,e.\ $\xi<1$, can be achieved in both the aggressive and the conservative approach.

\begin{figure}
\centering
\includegraphics[width=.475\textwidth]{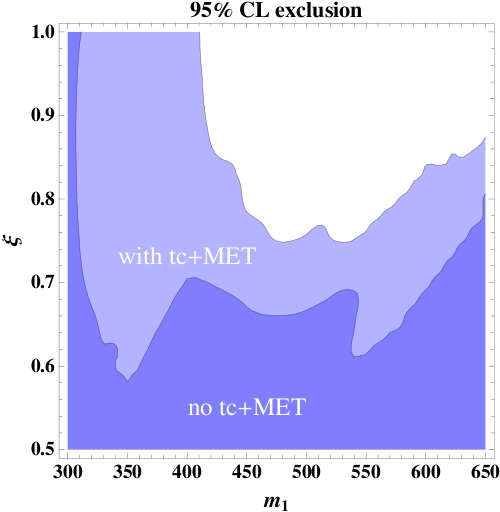}
\caption{Exclusion contours in the $(m_1,\xi)$ plane \cite{Blanke:2013uia}, where $m_1$ is the stop-like mass and $\xi$ is the naturalness parameter defined in eq.\ \eqref{eq:xi}. The dark (light) shaded area corresponds to the 95\% C.L. excluded region obtained in the aggressive (conservative) approach.  \label{fig:naturalness}}
\end{figure}

To study the improvement in naturalness and its correlation with the stop-like mass more explicitly, Figure~\ref{fig:naturalness} shows the 95\%\,C.L. exclusion contours  in the $(m_1,\xi)$ plane. Again we observe both in the aggressive and the conservative approach that stop-scharm mixing allows to significantly soften the lower bound on the stop-like mass $m_1$, with the improvement being stronger in the aggressive estimate. In addition the lowest values for $\xi$, corresponding to the least fine-tuning in $m_{H_u}^2$, can be obtained for stop-like masses around $450-500\,\text{GeV}$ in the conservative approach and $350\,\text{GeV}$ in the aggressive approach. Recall that these two approaches should be considered as the two extreme cases when trying to estimate the effects of the $t\bar c(c\bar t)+\EmissT$ final state on $\text{jets}+\EmissT$ searches, and that the precise situation is expected to lie in between these two scenarios.

Before moving on it is instructive to assess the impact of a non-zero neutralino mass on the bounds discussed above. For small LSP masses $\lsim 150\gev$ the experimental bounds on the squark masses are almost independent of the value of the LSP mass, hence in this case we expect a very small impact on our analysis. For larger LSP masses the on average smaller $p_T$ of the decay products becomes relevant, resulting in weaker constraints on the squark masses. The most important effect of finite LSP mass is the difference in phase space between the $t\chi_1^0$ and $c \chi_1^0$ final states, which becomes relevant if $m_t + m_{\chi_1^0}\lsim m_{1,2}$. This leads to a suppression of the $t\chi_1^0$ final state with respect to the $c\chi_1^0$ one. Consequently since the bound on the $\text{jets}+\EmissT$ final state is weaker than the one on $t\bar t+\EmissT$, the phase space suppression is beneficial for lowering the mass bounds on the mixed $\tilde c_R -\tilde t_R$ states.

\section{Possible signatures}

\begin{figure}
\centering
\includegraphics[width=.45\textwidth]{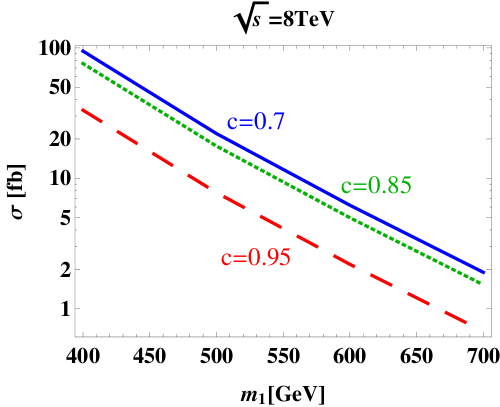}
\caption{LO cross-section prediction \cite{Blanke:2013uia} for the $t\bar c (c\bar t)+\EmissT$ for various values of the mixing parameter $c=\cos\theta_R^{ct}$ at a centre of mass energy $\sqrt{s}=8\,\text{TeV}$. \label{fig:tcMET8}}
\end{figure}

\begin{figure}
\centering
\includegraphics[width=.45\textwidth]{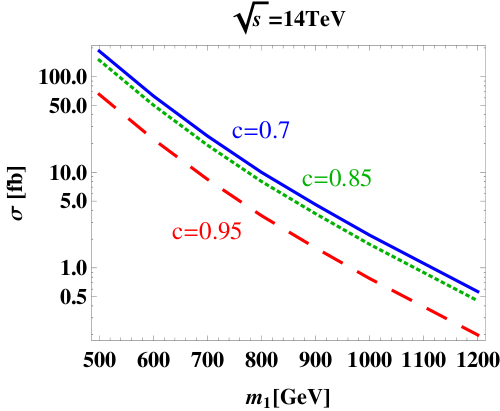}
\caption{The same as Figure \ref{fig:tcMET8}, but for $\sqrt{s}=14\,\text{TeV}$. \label{fig:tcMET14}}
\end{figure}

After estimating the effect of flavour violation on the current squark mass bounds, we move on to discuss signatures of a large mixing between the right-handed stop and scharm. One possibility has already been identified in the above discussion of the current LHC bounds: The misalignment between stop and scharm flavour and mass eigenstates leads to the flavour violating final state $t\bar c(c \bar t)+\EmissT$ that can successfully be searched for already in the present $8\,\text{TeV}$ data set. Figure \ref{fig:tcMET8} shows the corresponding leading order (LO) cross-section as a function of the lighter squark mass $m_1$, for various values of the mixing parameter $c$. A dedicated search for this signature, in the absence of signal, would already place significant limits on the parameter space of stop-scharm mixing, and we encourage the experimental collaborations to analyse this channel. At the 14\,TeV LHC it will be possible to probe this channel for squark masses even beyond 1\,TeV provided the mixing angle is large, see Figure \ref{fig:tcMET14}.

In principle stop-scharm mixing also leads to the production of same-sign tops at the LHC: the signature $tt+\EmissT$ and $\bar t \bar t+\EmissT$ are generated with the same rate. The underlying process is a $t$-channel gluino exchange between two charm or anti-charm quarks, with the pair-produced same-sign squarks both decaying into $t+\chi^0_1$. In practice however this process is suppressed by the small charm quark PDF, by a flavour mixing factor $c^4 s^4\le 1/16$ and by the requirement that both tops decay semileptonically. Consequently the cross section for this signal turns out to be too small to be observable even with $300\,\text{fb}^{-1}$ of 14\,TeV data and a luminosity upgrade is needed for a possible observation.

Last but not least we mention that stop-scharm mixing in the right-handed sector also gives rise to interesting effects in flavour changing neutral current observables. In particular a sizeable CP asymmetry in $D^0\to K^+K^-/\pi^+\pi^-$ decays could be generated by a large stop-scharm mixing angle, if accompanied by a large scalar coupling $A_t$.

\section{Conclusions}

Supersymmetry, in order to naturally explain the scale of electroweak symmetry breaking, requires the existence of light scalar top quarks, the stops, below the TeV scale. While a lot of effort has been put into experimental searches for these particles, no hint for their existence has yet been found and the limits on their mass become increasingly stringent.
In order to fully exploit the parameter space describing the relevant production and decay modes, it is important to go beyond the simplest models in which a 100\% branching ratio into $t\chi^0_1$ (or $b\chi^+_1$) is realised. 

Our analysis \cite{Blanke:2013uia} considers a particular extension of this simple scenario: By allowing for a large flavour mixing angle between the right-handed stop and scharm, a new final state $c\chi_1^0$ opens up and effectively suppresses the $t\bar t+\EmissT$ final state. At the same time, the stop-like state can be ``hidden'' in the $\text{jets}+\EmissT$ channel, thanks to the much larger SM background. Consequently the bound on the stop-like mass becomes significantly weaker, while the bound on the scharm-like mass remains mostly unaffected. The weaker mass bounds then also manifest themselves in a modest improvement of naturalness in the EW sector.

After the completion of our analysis more stringent constraints on stops decaying into $t\bar t+\EmissT$ became available \cite{stop-search-status}, reaching up to around $650\gev$ for a massless neutralino. While our numerical results will of course be affected by the inclusion of these additional data, the stop mass bound can always by weakened by stop-scharm mixing as long as the scharm is less stringently constrained than the stop.  Since the pair production cross-section is equal for both states and the backgrounds for $\text{jets}+\EmissT$ is much larger than for $t\bar t+\EmissT$, we do not expect this pattern to change.

A large stop-scharm flavour mixing predicts the flavour violating final state $t\bar c(c\bar t)+\EmissT$ to be produced with a significant cross-section. A dedicated experimental search appears very promising to either discover or severely constrain this set-up already with the present 8\,TeV data set.

\begin{acknowledgement}
I would like to thank Gian F.\ Giudice, Paride Paradisi, Gilad Perez and Jure Zupan for a very pleasant and inspiring collaboration that led to the results presented here and published in~\cite{Blanke:2013uia}. 
\end{acknowledgement}

%
%
%

\end{document}